\documentclass{vgtc}                          

\ifpdf
  \pdfoutput=1\relax                   
  \usepackage{graphicx}                
  \usepackage{soul}
  \DeclareGraphicsExtensions{.pdf,.png,.jpg,.jpeg} 
\else
  \usepackage{graphicx}                
  \DeclareGraphicsExtensions{.eps}     
\fi%

\graphicspath{{figures/}{pictures/}{images/}{./}} 

\usepackage{microtype}                 
\PassOptionsToPackage{warn}{textcomp}  
\usepackage{textcomp}                  
\usepackage{mathptmx}                  
\usepackage{times}                     
\usepackage{cite}                      
\usepackage{tabu}                      
\usepackage{booktabs}                  
\usepackage{tikz}
\usepackage{tabularx}
\usepackage{enumitem}

\tikzset{cross/.style={cross out, draw, minimum size=2*(#1-\pgflinewidth), inner sep=0pt, outer sep=0pt}}

\usepackage{amssymb}
\usepackage{pifont}
\newcommand{\cmark}{\ding{51}}%
\newcommand{\xmark}{\ding{55}}%

\usepackage{xcolor}
\newcommand*\circled[1]{\tikz[baseline=(char.base)]{
            \node[shape=circle,fill,inner sep=1pt] (char) {\textcolor{white}{#1}};}}

\onlineid{5775}

\vgtccategory{Research}

\vgtcinsertpkg

\title{FiberStars: Visual Comparison of Diffusion Tractography Data between Multiple Subjects}

\author{Loraine Franke\thanks{e-mail: franke@mpsych.org}\\ %
        \scriptsize University of Massachusetts, Boston %
\and Daniel Karl I. Weidele\thanks{e-mail: daniel.karl@ibm.com}\\ %
     \scriptsize IBM Research %
\and Fan Zhang \thanks{e-mail: fzhang@bwh.harvard.edu}\\ %
     \scriptsize Harvard Medical School %
\and Suheyla Cetin-Karayumak\thanks{e-mail: skarayumak@bwh.harvard.edu}\\ %
     \scriptsize Harvard Medical School %
\and Steve Pieper \thanks{e-mail:  pieper@isomics.com}\\ %
     \scriptsize Isomics, Inc. %
\and Lauren J. O'Donnell\thanks{e-mail: odonnell@bwh.harvard.edu}\\ %
     \scriptsize Harvard Medical School %
\and Yogesh Rathi \thanks{e-mail: yogesh@bwh.harvard.edu}\\ %
     \scriptsize Harvard Medical School %
\and Daniel Haehn\thanks{e-mail: haehn@mpsych.org}\\ %
    \scriptsize University of Massachusetts, Boston %
    }

\teaser{
 \includegraphics[width=0.9\linewidth]{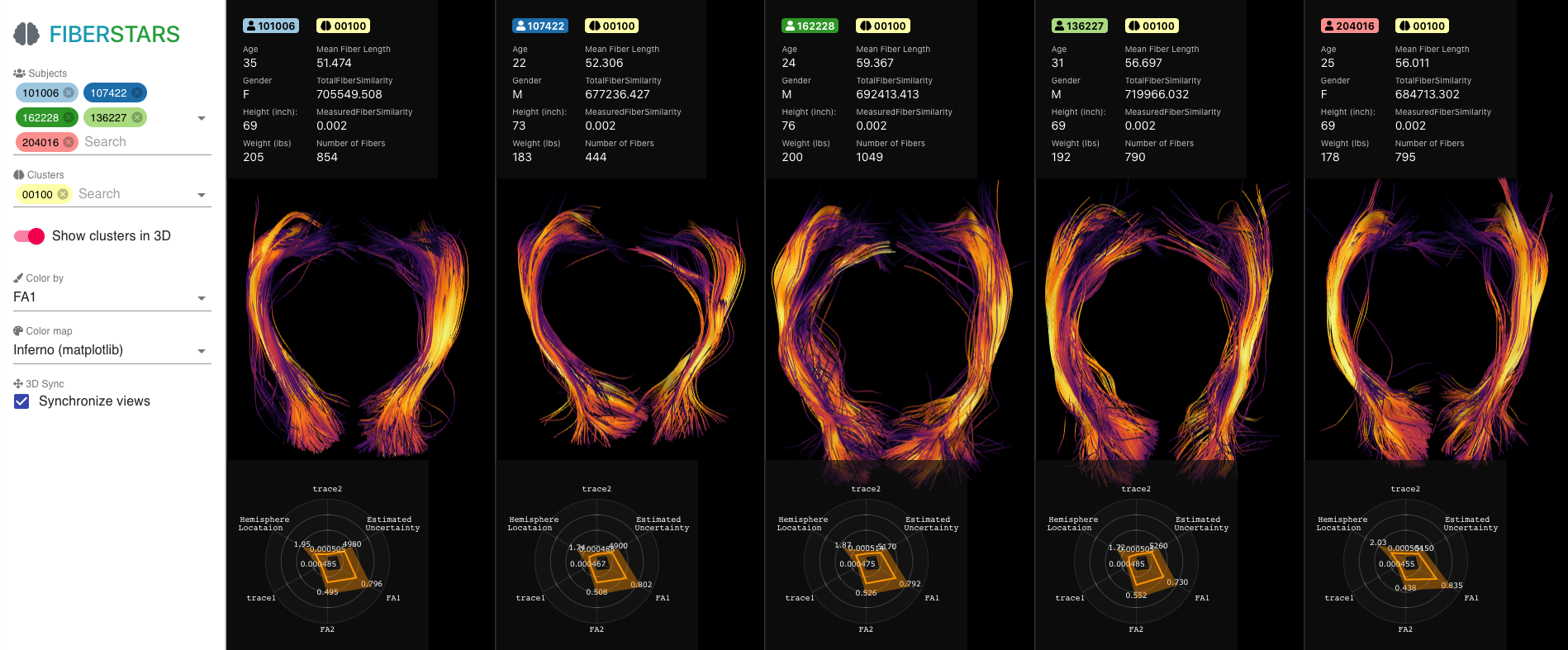}
 \caption{\textbf{Comparison of fiber clusters across subjects in split screen mode of FiberStars.}
 Human Connectome Project
tractography data from six different subjects. The bottom two-dimensional mappings allow for a faster comparison. A demo of FiberStars
with a small data sample can be found at https://lorifranke.github.io/ABCD/}
  \label{fig:teaser}
}

\abstract{
Tractography from high-dimensional diffusion magnetic resonance imaging (dMRI) data allows brain's structural connectivity analysis. Recent dMRI studies aim to compare connectivity patterns across subject groups and disease populations to understand subtle abnormalities in the brain's white matter connectivity and distributions of biologically sensitive dMRI derived metrics. Existing software products focus solely on the anatomy, are not intuitive or restrict the comparison of multiple subjects. In this paper, we present the design and implementation of FiberStars, a visual analysis tool for tractography data that allows the interactive visualization of brain fiber clusters combining existing 3D anatomy with compact 2D visualizations. With FiberStars, researchers can analyze and compare multiple subjects in large collections of brain fibers using different views. To evaluate the usability of our software, we performed a quantitative user study. We asked domain experts and non-experts to find patterns in a tractography dataset with either FiberStars or an existing dMRI exploration tool. Our results show that participants using FiberStars can navigate extensive collections of tractography faster and more accurately. 
All our research, software, and results are available openly.

} 

\CCScatlist{
  \CCScatTwelve {Neuroscience} {Brain Fibers} {Diffusion MRI} {High-Dimensional Data} {3D Visu\-al\-iza\-tion} {Visu\-al\-iza\-tion techniques}
  }

\firstsection{Introduction}

\begin{document}

\firstsection{Introduction}
\maketitle

In recent years, studying the brain and its neural connectivity has become an emerging discipline among various research fields. Especially, \textit{diffusion magnetic resonance imaging} (dMRI) is currently the only technique that enables tracing the structural anatomy of white matter tracts in-vivo in the human brain. DMRI is sensitive to molecular water diffusion, charaterizes subtle changes in the brain microstructure and measures structural connectivity abnormalities in white matter tracts~\cite{basser1994estimation}. To analyze the white matter tracts and construct maps or diagrams of the brain's structural connectivity with high-resolution images, researchers use a process called tractography~\cite{basser2000vivo,lichtman2011big, conturo1999tracking}. DMRI tractography has gained in popularity in clinical practice and research on brain diseases such as autism, multiple sclerosis, stroke, dementia, and schizophrenia~\cite{assaf2008diffusion, thomason2011diffusion}. Moreover, dMRI is a powerful tool to track and detect disruptions in structural connectivity regarding brain disease and disorders \cite{fornito2015connectomics, gori2016parsimonious}. For example, when comparing the brain connectivity between healthy and disease populations, it is critical to understand the potential pathology.

Tractography data needs interpretation to be useful, and therefore visualizations are required to understand the underlying tissue microstructure of fiber tracts. High-dimensional fiber tracking datasets consisting of tens of gigabytes in size with millions of fibers, and the spatial 3D characteristics yield fundamental challenges for data exploration and visualization. Tractography data can include millions of 3D polylines with each line representing the path of a single white matter tract. These lines can form a fiber bundle or also called a cluster. The data is highly variable across fiber cluster and subjects.

\paragraph{\textbf{Goals.}} Our main goal is to visualize fiber data in an efficient way that allows comparisons between different clusters or bundles and subjects. We choose to pair existing 3D visualizations with a 2D approach to reduce the complexity of the data. FiberStars aims to assist researchers among various disciplines, including neuroscientists, neurosurgeons, and psychologists. State-of-the-art tools used by clinicians are linked to an extensive workload and explicitly require to navigate through massive amounts of data independently. We design a web-based analysis tool for brain connectivity research that is easy to use for novices and experts alike. Another goal is that users do not necessarily require a detailed understanding of complex relationships and patterns in the data. With FiberStars, users can generate and validate new hypothesis when comparing tractography of multiple subjects with several levels of abstraction. We visualize large multi-subject datasets in a projection view that shows the overall distribution. A compact 2D representation creates fingerprints for different subjects and fiber clusters. Finally, we support the paired visualization of 3D anatomical renderings with 2D representations across multiple subjects and multiple regions of interest across different devices.
We build off existing visualization research, that has demonstrated effectiveness of additional two-dimensional representations in medical imaging, such as for connectomics \cite{al2014neurolines} or other fields such as cerebral arteries \cite{pandey2019cerebrovis}.

\paragraph{\textbf{Contributions.}} While dMRI is a very specific use case, it is representative of a large class of complex visualization challenges involving multidimensional data or data composed of collections of subjects, and each with multiple 3D shapes with spatially-varying properties. Other contexts yielding similar visualization challenges are, for example, the visualization of biological diversity measured with microCT or surface scanning of specimens, or the comparison of large multidimensional astronomical data collected with multi-spectral telescopes. Overall, increasing amounts of these types of data are being collected in a variety of fields, while most of the current visualization methods are still insufficient.

Therefore, we present the design, implementation, and evaluation of FiberStars for the specific use case of dMRI data. However, we hope to inspire and contribute to other fields with similar data and visualization challenges. Our application facilitates the analysis of high-dimensional diffusion MRI data with different levels of abstraction. We focus specifically on ensemble visualization to allow the direct comparison of regions of interest and across multiple subjects. Such comparisons are important as tractography datasets are getting larger and include multiple timestamps. FiberStars maximizes usability, and our quantitative user study shows that novices without any tractography experience can generate meaningful insights. We also evaluate FiberStars with tractography experts and show that our software allows faster and more precise analysis compared to alternative state-of-the-art tools. All our materials and software with documentation are openly available on GitHub at \url{https://github.com/lorifranke/FiberStars}.

\section{Related Work}
\label{sec:related_work}
Among various scientific disciplines, the development of interactive three-dimensional renderings plays an increasingly important role in data and information visualization. In recent years, the field of neuronal connectivity visualization of brain imaging data has emerged. Current tools and libraries such as XTK~\cite{haehn2014neuroimaging}  and Fiberweb~\cite{ledoux2017fiberweb} contribute to this task and allow web-based 3D renderings. However, such visualizations can be complex and hard to understand. Many works present approaches to visually explore the complex topology of biological datasets~\cite{jianu2011exploring,jianu2009exploring,poco2012employing,chen2009novel}. To further decrease the cognitive load for visualization consumers, researchers suggest visualizing high-dimensional data with a reduced representation for data exploration and analysis~\cite{borkin2011evaluation}.

Neurolines~\cite{al2014neurolines} also provides dimensionality reduction. Researchers here visualize 3D brain tissue data as a 2D subway map. This multi-scale approach allows scientists to study connectivity with much greater ease than working in 3D. Further, Mohammed et al.~\cite{mohammed2017abstractocyte} visualize similar datasets with different levels of detail. 
Another example is Jianu et al.~\cite{jianu2009exploring}, where abstract 2D paths represent brain fibers while preserving anatomical information. 
Further literature aims to automatically cluster fibers providing similarity measures among fibers or whole fiber bundles. The remaining need for exploring fiber bundles was approached by~\cite{chen2008abstractive, cauteruccio2015automated, kamali2016automated, chen2015fiber, zhang2018anatomically}. But all these visualizations cannot analyze multiple datapoints across different subjects. Other related research has shown comparative visualizations, for example, using fMRI and multivariate clinical data~\cite{jonsson2019visual} with parallel coordinates or on tenser changes of DTI with scatterplots~\cite{abbasloo2019interactive}. 

However, to allow comparative visualizations of DTI data, researchers need software that allows cohort and ensemble visualizations. One example is DiffRadar by Mei et al.~\cite{mei2016visually}, a combined 2D dimensionality reduction with 3D fiber visualization using multidimensional scaling (MDS). Yet, to allow MDS, all fibers need to have the same number of vertices, and the data requires resampling, which might lead to distortion. 

Another available application was created by Yeatman et al.~\cite{yeatman2018browser}. The authors propose the AFQ-Browser (Automated Fiber Quantification) related to another previous approach called BundleMap~\cite{khatami2017bundlemap}. This tool enables quantitative analysis of white matter fiber tracts and comparisons across different subjects. The authors compare healthy subjects with subject groups suffering from Multiple sclerosis. In initial experiments of using AFQ-Browser on our data, we faced challenges for effective comparisons across subjects, which will be discussed in Section 5. We design FiberStars to overcome these limitations and carefully compare our software to the AFQ-Browser in this paper. In Section \ref{sec:study}, we evaluate the limitations of each application in more detail.


\section{Scientific Background}

DMRI by Basser et al.~\cite{basser1994estimation} allows exploring information from in vivo fibrous structures such as white matter or muscles and is widely used across hospitals, universities, and research centers. In collaboration with neuroscience researchers, we studied how we can efficiently visualize dMRI data of single and multiple subjects.

Tractography data represents trajectories of fibers (or streamlines) in the white matter. These paths are positional data with different series of vertices ($x,y,z$) in 3D. Researchers then often attach per-vertex scalars or per-fiber properties to include additional information such as acquisition parameters, diffusion properties, or quantities obtained during processing or analysis. 
Researchers use this information to estimate cellularity, size of cell bodies and processes, or presence of myelin during the diffusion process~\cite{avram2016clinical}.


\paragraph{\textbf{ADHD dataset.}} We tested our tool on different datasets. The first dataset contains dMRI scans of subjects suffering from attention deficit hyperactivity disorder (ADHD)~\cite{zhang2018suprathreshold, rathi2011sparse, wu2019detecting}. High-resolution MR images were obtained on a Siemens 3T scanner at Boston Children's Hospital, Boston, USA, with approval of the local ethics board. Multi-shell diffusion-weighted imaging (DWI) data were acquired using a simultaneous multi-slice acquisition factor of 2 at a spatial resolution of 2×2×2 mm$^3$ with 70 gradient directions spread over the three b-value shells of 1000/2000/3000 s/mm$^2$. Whole-brain tractography was conducted using the unscented Kalman filter tractography (UKF) method, from the ukftractography 
package~\cite{reddy2016joint, 5559623, rathi2014multi}. During fiber tracking, the following scalars were recorded, including the normalized signal estimation error, signal means, return-to-origin probability (RTOP), return-to-plane probability (RTPP), and the return-to-axis probability (RTAP). 
The ADHD dataset includes 67 subjects, each containing 800 clusters.
\paragraph{\textbf{HCP dataset.}} Our second dataset is from the Human Connectome Project (HCP)~\cite{van2013wu}. HCP data was acquired with a customized Connectome Siemens Skyra scanner, and acquisition parameters TE = 89.5 ms, TR = 5520 ms, phase partial Fourier = 6/8, and voxel size = 1.25 x 1.25 x 1.25 mm$^3$. A total of 288 images were acquired for each subject, including 18 baseline images with low diffusion weighting b = 5 s/mm$^2$ and 270 diffusion weighted images evenly distributed at three shells of b = 1000/2000/3000 s/mm$^2$.
Scalars include diffusion measures of the fractional anisotropy (FA), mean diffusivity (MD), and the hemisphere location. Changes in these diffusion scalars are considered to reflect alternations of the underlying tissue microstructures. Quantifying changes is helpful for monitoring disease and abnormalities, for example, inflammation, cell death, changes in myelination, edema, gliosis, increase in connectivity of crossing fibers or in extra- or intracellular water and many more~\cite{o2015does,basser2011microstructural}. Besides scalars, each fiber bundle contains cell data with properties such as embedding coordinate, cluster number, embedding color, total fiber similarity, and measured fiber similarity. Especially, total and measured fiber similarity in a fiber tract are of special interest in terms of crossing fiber bundle comparisons. Interpretation of changes in those scalar measurements is a complex task due to their non-specificity.  Our tool supports any type and number of scalars attached to the fiber data. The data is available in .TRK- and .VTP-file formats, which are XML type files. Both datasets include data of the right and left hemisphere of the brain. Each file includes data for one fiber bundle. 
The HCP dataset includes 100 subjects and additional metadata (such as patient demographics) as CSV-files. 

\subsection{Domain Goals and Design Requirements}
In regular meetings with our collaborating scientists, we discussed goals and possible visualization designs. Through semi-structured interviews, we explored which type of visualizations are helpful for domain specific tasks. As most of the recent works show limitations in terms of multi-subject-multi-cluster comparisons, we decided to develop FiberStars. We derived the following requirements for a visualization tool: 
{\textbf{(R1) Multi-resolution: Multi-level visualizations from individual fiber clusters to whole-brain analysis.}} Most existing work focuses on single-fiber visualization only, which is useful for surgical planning and individual diagnosis in a clinical environment. Besides retrieving information of a certain fiber bundle, the system should support direct comparisons on the level of multiple fiber bundles of a subject's brain. 
\textbf{(R2) Allowing to compare between multiple subjects.} Our collaborators explicitly need group comparisons to study longitudinal scans from healthy subjects and those who suffer from diseases. The original purpose of FiberStars was to use the software for the Adolescent Brain Cognitive Development (ABCD) study. ABCD is the largest long-term study of brain development and child health in the United States~\cite{jernigan2018adolescent}. With a targeted visualization tool, we can associate levels of brain development between subjects and co-founding factors such as water quality, pollution, social and lifestyle behaviors, physical activity, and others. Therefore, our domain experts need to assess anatomical structures, and the system should provide three-dimensional views paired with 2D views to make white matter tract differences in tractography easier and faster to identify. \textbf{(R3) Interpretability and Usability: Provide an intuitive visual design that allows fast comparison and detection of group patterns, outliers or abnormalities in the high dimensional data.} Even without prior knowledge of patterns in the data, the user can gain new insight into the dataset and identify subjects or clusters that are different from others depending on the properties or scalars. This can help domain experts to explore data faster and more easily. The targeted visualization tool requires scalable, interactive elements that allow working with high-dimensional data. 

\paragraph{\textbf{Usage Scenario and Task Abstraction}}\label{sec:usageabstraction}

Our domain experts work with extensive dMRI data collections that require a complex computational infrastructure for processing, storage, and visualization. FiberStars provides a web-based user interface to explore, view and query the data for a user-defined set of subjects from an entire cohort of subjects across different diagnostic categories, including anatomical 3D visualizations. Our researchers perform data quality control in collections of unprecedented size, test hypotheses and answer research questions by comparing fiber tract changes in longitudinal studies for brain development as presented in R2. Clinical studies include the analysis of factors combined with a pathological finding, e.g. the fractional anisotropy (FA) in the corpus callosum in patients with auditory verbal hallucinations is reduced compared to the controls~\cite{dibiase}.
We derived multi-level tasks (\textbf{R1}/\textbf{R2}) with the task taxonomy of Brehmer and Munzner \cite{brehmer2013multi}:
\textbf{T1: Analyze an abnormal measurement.} When a possible outlier or abnormal cluster of a subject is pre-selected or has been identified in a previous step, the user might want to investigate how this measurement contributes to its abnormality. For example, we check if other measurements/scalars are abnormal or identify the abnormality's location along the fiber tract by browsing for deviations. T1 is derived from the interpretability and usability in \textbf{R3} testing both 2D and 3D representations in each tool for the simple case of a single subject single cluster. The user is asked to browse and compare the relevant measurement as input to get an anomaly as output.
\textbf{T2/T3: Comparing.} With relevant measurements as input, the user explores, compares and identifies~\cite{brehmer2013multi} a certain measurement, or scalar value. Users can test prior assumptions and hypotheses about the selected cluster or selected subject of interest. In \textbf{T2} the user has previously identified a subject of interest with anomalies, and then compares this previously identified cluster to multiple, or all other clusters, in a single subject's brain (\textbf{R1}). Vice versa in \textbf{T3}, a possible scenario is to compare an identified cluster to the same cluster of the other apparently healthy subjects (\textbf{R2}). 
\textbf{T4: Identify anomalies/outliers and extreme values among multiple subjects and clusters.} For an in-depth analysis, the user needs to discover, explore, and identify a relevant cluster or subject in a full collection (\textbf{R2}). Additionally, the user wants to retrieve information, such as the subject's metadata. According to~\cite{brehmer2013multi} the user should discover, explore and identify one relevant cluster within the input of all available clusters. 
\textbf{T5: Interaction within a cluster.} Currently, our domain experts are unable to quantify differences in tract shape automatically. With FiberStars' 3D visualization and different color schemes, the user can interactively analyze the dMRI data. \textbf{T5} measures overall performance and usability (\textbf{R3}). The user is asked to navigate, select and change a given anomaly, attribute or feature with a different color or cluster as output.

\section{Visualization Design}

This sections describes the final design choices of FiberStars.

\subsection{Visual Elements and Design Choices}\label{sec:design}
To tackle the challenge of creating a customized visualization software for the exploration process, we conceptualized FiberStars in close collaboration between neuroscientists and visualization researchers working with DTI data in an iterative process. Our collaborators regularly work with large-scale tractography data. The following five visual components offer tractography exploration on different levels and controlled navigation of dMRI data collections:


\paragraph{\textbf{Universal Toolbar.}}

The users can navigate through different views by selecting and deselecting several options in the navigation toolbar on the left. In the first place, it is possible to select the desired subjects and clusters. When selected one or multiple subjects, the fiber tracts are projected in the 2-dimensional projection view, and the navigation toolbar opens additional alternatives to select (\textbf{R1}). The user can select the scalar of interest and has different options on how to color the points in the projection view (see Section Projection View). The user can easily switch between multi-cluster view and split-screen view by enabling the 3D slider button (\textbf{R3}). Furthermore, the navigation toolbar offers different ways to color the 3D fiber tracts with dropdown menus.

\paragraph{\textbf{Projection View.}}

\begin{figure}
  \centering
  \includegraphics[width=\linewidth]{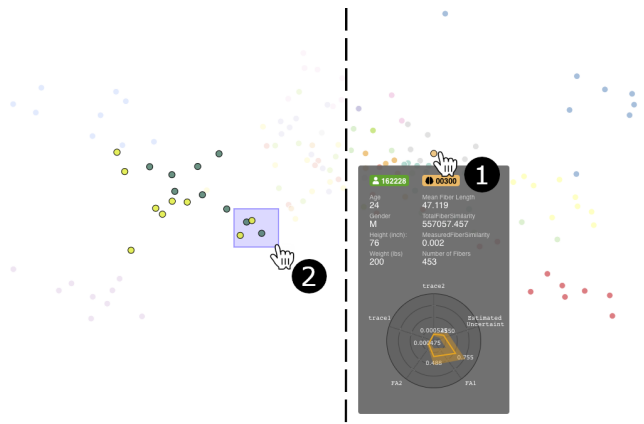}
  \caption{\textbf{Projection view shows distribution of multiple clusters and subjects.} When hovering over a point in the projection view, the user can get more detailed information on the projected cluster (1). 
  The Projection View allows the user to select interesting clusters by clicking and dragging over a group of multiple points in the projection view. This also highlights the same cluster in other subjects (2).}
  \label{fig:projection_view}
\end{figure}

The projection view (Fig. \ref{fig:projection_view}) of FiberStars serves as an entry point for users without in-depth a priori knowledge about clusters. In the Universal Toolbar users can select subjects $s \in S$ from the data collection. As we currently only support manual sampling of subjects, automated or guided approaches are subject to future work. Only after drawing a subject from the collection, its associated clusters $C_s$ are loaded lazily into FiberStars: $C = C_1 \cup C_2 \cup ... \cup C_{|S|}$. Then, for every cluster $c \in C$, we seek layout coordinates $x_c \in \mathbb{R}^{d}$ in the Euclidean plane ($d=2$). In particular, we would like to preserve distances $\delta(c_i, c_j) \approx ||x_{c_i} - x_{c_j}||$ with $c_i, c_j \in C$ and  $\delta: C \times C \rightarrow \mathbb{R}$ a distance function operating in the high-dimensional space of cluster scalars. In our application we find this problem formulation, known as \emph{Multidimensional Scaling} (MDS)~\cite{torgerson1952multidimensional}, to be favorable over alternative dimensionality reduction techniques~\cite{mcinnes2018umap, maaten2008visualizing}.
Firstly, a less distorting approach allows for more intuitive reasoning, especially when expanding the exploration process from smaller to larger numbers of scalars (\textbf{R1/R2}). For example, consider a domain expert primarily interested in the relationship between the following scalars: Fractional Anisotropy (FA) and Estimated Uncertainty (EU). Then, the MDS solution in the projection view reduces to a simple, reasonable scatter plot-like layout to begin with. The domain expert can use the universal toolbar to add or replace scalars of interest to/from the scatter plot, which allows forming an intuition on the impact of the different variables. Further guidance in the resulting layout could be provided by adding artificial data points, which can be obtained by fully maximizing a single scalar while fixing all others to the minimum~\cite{weidele2016graphical}. Secondly, we find that an analytical solution to the dimensionality reduction problem tends to be more user-friendly in that it is robust and requires no fine-tuning of artefactual parameters (\textbf{R3}). In FiberStars, our method of choice is PivotMDS~\cite{brandes2006eigensolver}, as it satisfies the above requirements and can efficiently scale to very large data sets. Mei et al.~\cite{mei2016visually} use a related technique, Landmark MDS~\cite{de2004sparse}, to compute projections for fibers \emph{within} a cluster. However, Brandes and Pich \cite{brandes2008experimental} show PivotMDS is superior to Landmark MDS in general graph layouts, a closely related problem. For better readability of individual data points in the already colorful space we further waive a density map overlay as suggested in \cite{mei2016visually}.

 Via the Universal Toolbar, the user can further map cluster scalar or subject attribute values to points in the projection. The technique allows the user to select multiple scalars as dimensions and set different colors for the individual points (e.g. color points by gender, age, cluster etc.) Moreover, the projection view hosts two action listeners (Fig. \ref{fig:projection_view}): \circled{1} Upon hovering a cluster point, a detailed pop-up summarizes statistics about the cluster, its corresponding subject and also displays the abstract cluster representation in the form of a radar chart that is further described in the next part of this section. \circled{2} Via rectangular brushing in the projection view, the user can choose clusters of interest, which will then be added to the alternative views across the application. Since we always display all selected clusters \emph{for all selected users} in the following views of the application, we highlight all these clusters in the projection, even if they are outside of the drawn selection window. This feature, we find, has a useful side-effect: corresponding clusters in other subjects can be more quickly identified (\textbf{R2/R3}). Thus, the distribution of these grouped clusters can be assessed directly within a potentially crowded point cloud.

\paragraph{\textbf{Abstract Representation.}} We choose consistent coloring of reappearing elements and information throughout FiberStars. For example, the same subject and the same cluster are always highlighted in the same color to avoid getting lost in the wealth of information. We implement color consistency in all views of the application. Furthermore, we leverage Pop-up elements that help the user to keep the orientation across different linked views (\textbf{R3}). In the projection view, we provide a Pop-Up when hovering over a single cluster point, as seen in \textbf{1} in Figure~\ref{fig:projection_view}, leading to more details of the cluster and the region of interest.

\begin{figure}
  \centering
  \includegraphics[width=\linewidth]{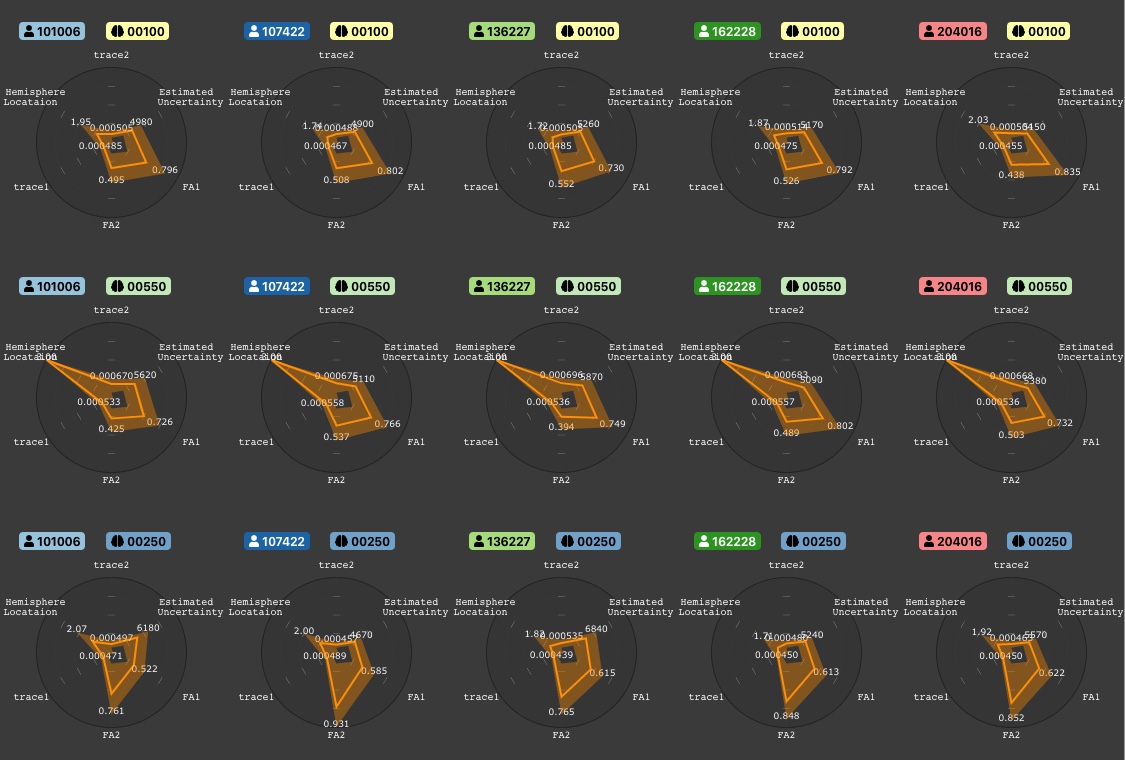}
  \caption{\textbf{Multi-Cluster view (\textit{Comparison Matrix}) of five different subjects and three different cluster.} The multi-cluster view allows to compare 2D radial plots visualizing multiple clusters of multiple subjects. It is useful for outlier detection of the mapped scalars and properties.}
  \label{fig:multicluster}
\end{figure}

\paragraph{\textbf{Comparison Matrix.}}
The Multi-Cluster view (or Comparison Matrix) in Figure~\ref{fig:multicluster} enables a comparison between multiple subjects and multiple fiber clusters. In this view, the user selects any number of subjects and clusters of interest in the navigation toolbar (\textbf{R2}). For each cluster, one 2D radial plot is shown with normalized scalars on each axis. Displaying multiple radial plots next to each other as in Figure \ref{fig:multicluster}, facilitates for the user to see the scalar differences between clusters. Especially, radial plots would allow a fast outlier detection of the scalars and properties, when multiple charts are shown next to each other. We evaluated alternative plots and designs for multi-dimensional data, for example line graphs as in~\cite{yeatman2018browser}, parallel coordinates similar to~\cite{jonsson2019visual}, as well as the option to integrate circular plots which we used in prototypes, heat maps or scatter plots. For exploratory purposes radial axes plots are useful to obtain an overview of the data~\cite{saary2008radar}. However, due to the nature of star-based visualizations being less visually complex for the eye, a user can find outliers or extreme values on the star axes faster when these are juxtaposed~\cite{draper2009survey}.
In the Multi-Cluster view other displayed information includes subject demographics and overall statistics such as mean fiber length, total fiber similarity, etc. When filtered by certain clusters and subjects of interest, the user can switch back any time to the 3D split-screen view (\textbf{R1}).

\paragraph{\textbf{3D Split-Screen Visualization.}} The split-screen view is shown in Figure~\ref{fig:teaser}. Here we integrate a typical 3D scientific view with additional information from the 2-dimensional radial plots. The Split-Screen view allows comparing fibers from multiple subjects side-by-side (\textbf{R2}). Camera interaction can be synced across all 3D views. Additionally, the users can choose between 120 different available color maps and color the 3D fibers by selecting a scalar of interest to map. Moreover, this enables finding areas along the fiber bundle with high or low scalar values, whereas the user can easily switch between the scalar of interest. Domain scientists want to verify assumptions or hypotheses by additionally consulting the 3D anatomy. With the 3D view, experts can locate exact areas along the fiber bundle containing anomalies or relevant measurements and draw conclusions for specific regions of interest in a subjects' brain. A typical use case of the final \textit{FiberStars} application can be seen in Figure~\ref{fig:teaser}. The three-dimensional fiber clusters in the background can be interactively explored. Multiple subjects are shown next to each other, allowing interpretation of 3D illustration with 2D data summaries.

\subsection{Implementation Details}
FiberStars builds interactive 2D and 3D visualizations with the JavaScript library. We use VTK.js for 3D renderings, a JavaScript implementation of the VTK software toolkit~\cite{schroeder2004visualization} which uses WebGL. For 2D visualizations, we included D3.js~\cite{bostock2011d3}. FiberStars is a Node.js application using React.js for the frontend. Adopting Node.js allows for flexible extensions in the future, such as additional visualizations, statistical plots, and other features. Being a web-based software, FiberStars runs hosted on a web server and does not require any client-side installation. For the first prototype, we brainstormed and experimented with different 2D charts to represent the scalars, where we grouped the data by scalar types and averaged the values of all scalars for all fibers. Then, we mapped these values to 2D using radial plots as described in Section~\ref{sec:design}. However, scaling was important. For example, in the ADHD dataset the scalar Normalized Signal Estimation Error has values between 0 and around 0.05, whereas the Estimated uncertainty has a range from negative to 22,000. We now perform Min-Max Feature scaling to normalize the ranges to allow comparison among subjects and clusters. 
React.js improves the overall usability and facilitates integration of additional components and features. With elements from the framework Material-UI, we were able to implement a modern user interface. This leads to the final design of FiberStars that seamlessly integrates different linked and interactive views that allow multi-dimensional data exploration. We support multiple tractography data formats. Our expert collaborators use VTK PolyData to store fiber clusters with a VTP file extension, offering a flexible data model and storing vertices, per-vertex scalars, and per-fiber properties as well as metadata without restrictions. VTK.js includes functionality to load fiber clusters from VTP files in JavaScript. Each cluster is then represented as VTK \textit{polydata}, describing a surface mesh structure that holds additional data arrays in points, cells, or in the dataset itself. After loading these arrays, we calculate the means of all assigned scalars and properties per fiber and cluster. However, Tractography clusters stored as VTP files can be in the order of hundreds of Megabytes, and a whole-brain tractography can be as large as multiple terabytes in size.



\paragraph{\textbf{Compressed Tractography Data.}}
Recently, we developed the Trako Compression Scheme~\cite{haehn2020trako}, and integrated it into FiberStars. Trako allows compressing .trk, .tck and .vtp files while achieving data reductions of over 28x. While Trako uses lossy compression, our experiments show no loss of statistical significance when replicating analysis from previously published tractography experiments. 
Paired with state-of-the-art 3D geometry compression algorithms, Trako allows fast data transfer and realtime visualization with nearly no preprocessing. As part of this paper, we present Trako file readers for the VTK.js, Three.js, and XTK~\cite{haehn2014neuroimaging} visualization frameworks. For FiberStars, we convert Trako files to VTK poly data structures.


\section{Comparative Evaluation Study}
\label{sec:study}

We evaluated the performance, effectiveness and usability of FiberStars within an extensive user study. With a between-subjects study design, we recruited non-experts (novices) and domain experts to compare our software to the existing state-of-the-art tool AFQ-Browser~\cite{yeatman2018browser}.
We performed an A/B comparison of both applications. The results of our study confirm our design decisions.
\paragraph{\textbf{Evaluation of Comparable Approaches.}}
\begin{figure}
  \centering
  \includegraphics[width=\linewidth]{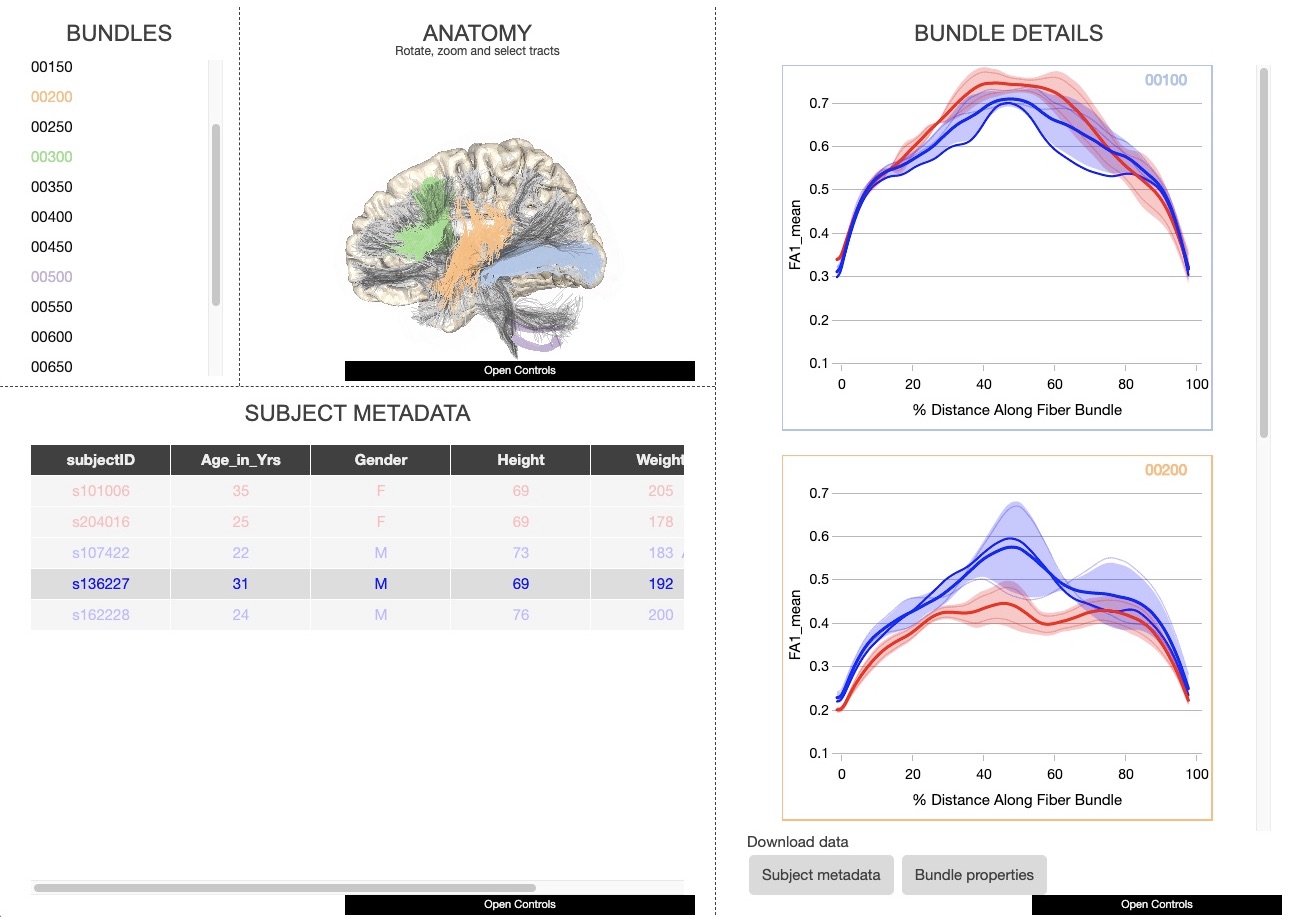}
  \caption{\textbf{User interface of AFQ-Browser.} Cluster selection in the list on the left and 3D anatomy view. The bundle details on the right allows comparing 2D representations of fibers. A metadata display on the left bottom provides more details about each fiber cluster.}
  \label{fig:afq}
\end{figure}

\begin{figure}
  \centering
  \includegraphics[width=\linewidth]{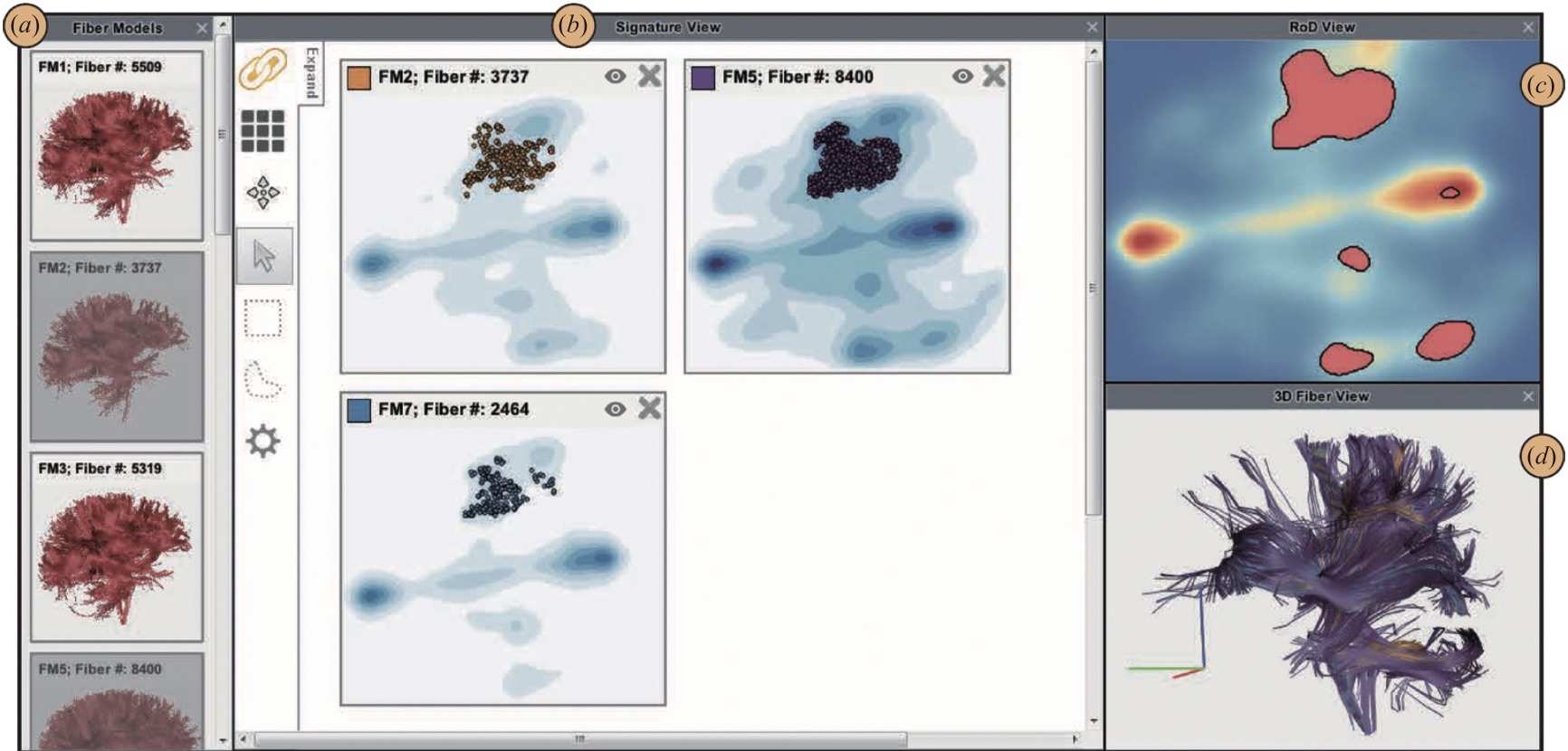}
  \caption{\textbf{User interface of Fiber Models (DiffRadar)} \cite{mei2016visually}. The user interface shows four different panels to get insights into a fiber cluster and single fibers. Left: Overview of each cluster. The center view shows more details into one selected cluster. Right top: shows regions of differences. Right bottom shows 3D model.}
  \label{fig:mei}
\end{figure}

\begin{table}[h]
	\centering
	\caption{Comparison of visualization tools with features that are offered in Fiber Models (DiffRadar) \cite{mei2016visually}, AFQ-Bowser \cite{yeatman2018browser}, and FiberStars (FS). FS and AFQ have most features in common.}
	\def\arraystretch{1.4}%
	\resizebox{\linewidth}{!}{%
	\begin{tabular}[t]{lccc}
		\toprule
		\textbf{Feature} & \textbf{Fiber Models} & \textbf{AFQ} & \textbf{FS (ours)}\\
		\midrule                        
		\textbf{Showing subject metadata}  & \xmark & \cmark & \cmark \\                              
		\textbf{Single fiber exploration}  & \cmark  & \xmark & \xmark\\                           
		\textbf{Select desired scalar or metric}  & \xmark & \cmark & \cmark\\                                
		\textbf{\vtop{\hbox{\strut Compare between fiber bundles}\hbox{\strut of individual subjects}} } & \xmark & in parts & \cmark\\
		\textbf{Interaction \& information retrieval of 3D model} & unknown & \xmark & \cmark\\
		\textbf{Visualization of large amounts of data} & unknown & in parts & \cmark \\
		\bottomrule
	\end{tabular}
	}
	\label{tab:toolcomparison}
\end{table} 
Most of the major work in this field requires previous knowledge of the patterns in the data; high customization for specific data of the lab or publication requires increased computational performance of the machine, or needs to be downloaded as local program. Further, related approaches mainly focus on exploring DTI data only on fiber level or on single subject level. 
Considering the most recent previous approaches, Yeatman et al.~\cite{yeatman2018browser} and Mei et al.~\cite{mei2016visually} are closest to our interactive tool. Table~\ref{tab:toolcomparison} includes the comparison of these tools and its most critical features for the exploration of dMRI data. Yeatman et al.'s AFQ-Browser has a user interface that consists of four different panels: a) bundles b) anatomy c) bundle details and d) subject metadata (Figure~\ref{fig:afq}).
While using their code, we noticed that transforming our dMRI files in the AFQ Browser only allows the input of a MatLab or JSON file format. Also, we were only able to use fibers with 100 data points. This might lead to a distortion of the data as fibers always have a varying number of points, and, therefore, we needed to interpolate the points. The other conventional tool, Fiber Models (DiffRadar) by Mei et al.~\cite{mei2016visually}, shows the differences between DTI fiber data by providing an \textit{intra}-cluster comparison of single fibers. Similarly to AFQ, the user interface in Figure \ref{fig:mei} consists of four panels.
The authors use a two-phase projection technique to map 3D fibers onto a 2D space with Multidimensional scaling (MDS). They implement a density estimation on their projection of the fibers. All fibers were reparameterized so that they have the same number of vertices and orientation by using the LAMP technique. This approach has already been shown in \cite{poco2012employing, chen2009novel}. The authors implement Landmark MDS (LMS) by randomly selecting some subset of fibers that are used as landmarks and then compute in a 2D plane the squared distance of fiber to the landmark fibers. 
A scatterplot is used to position similar fibers in a cluster next to each other, where they apply a Kernel density estimation to produce a continuous 2D density map of the scatterplot.
Unfortunately, we were not able to include this software in our quantitative user study. Nevertheless, this approach might only allow intra-cluster comparison of single fibers but no possibility among multiple subjects and clusters.


\paragraph{\textbf{Hypotheses.}}

We propose three hypotheses to validate the design of FiberStars:

\textbf{H1: FiberStars provides higher usability than AFQ-Browser.} Both user groups evaluate the usability of each tool and how they perceive working with it. During the development of FiberStars, we focused on overall usability and intuitiveness with minimal and slick user experience in mind. We predict that participants using FiberStars report higher usability than the ones working with AFQ-Browser.

\textbf{H2: Analyzing DTI data is more accurate in FiberStars.} The users in our study are presented with the same datasets in both tools. FiberStars is optimized towards analyzing and comparing tractography data of multiple subjects. With matching amounts of training, we predict that novices and domain experts will more accurately explore scalars and properties of fiber clusters with FiberStars than with AFQ-Browser. We measure accuracy in terms of correct answers for the tasks.

\textbf{H3: Within a given timeframe, the users are faster in solving tasks with FiberStars than with the alternative tool.} We measure the time it takes participants to complete pre-defined tasks. These tasks were defined in connection with diffusion imaging researchers and replicate day-to-day use-cases of domain experts. Since we designed the FiberStars application with constant feedback and input by domain experts, we predict that participants using our software perform more efficiently.

\paragraph{\textbf{Participants.}}
At first, we evaluate the usability of both tools FiberStars and AFQ-Browser with participants without prior knowledge of DTI or tractography, recruited through flyers and mailing lists. We estimated to recruit 11 participants per tool including a dropout rate of roughly 10\%~\cite{faulkner2003beyond, hwang2010number, nielsen1993mathematical}. From initial 22 participants, we had to exclude 2 subjects due to technical issues during the online meeting. 
Twenty participants completed the full study ($N=20$). 13 participants were females and 7 males, with an age range from 18 to 39 years, consisting of students and workers with a variety of backgrounds. All participants reported not having any visual impairments. Participants received monetary compensation for their time. Additionally, we asked 6 domain experts to participate in our user study, testing one of the two tools. Qualifying domain experts are researchers who perform complex data analyses with dMRI or DTI data who have not used either AFQ-Browser or FiberStars before. We also excluded all researchers that helped during the design of FiberStars. 

\paragraph{\textbf{Data.}} 
For fair evaluation and comparativeness, we used the same data in both tools. We randomly picked a sample of 5 subjects from the Human Connectome project and to further reduce data loading times, selected every 50\textsuperscript{th} of the 800 clusters (total 16 clusters). We added the corresponding metadata to both tools, including information such as subject ID, age, gender, weight, and height.


\paragraph{\textbf{Tasks.}}

Following Section~\ref{sec:usageabstraction}, we derived concrete tasks evaluating both tools in a controlled experiment. An expert user could get insights into the data after having identified a specific abnormal cluster (T1), or similarly, compare clusters from tractography of a single diseased subject (T2). T3 helps to identify a subject having anomalies in certain scalar values. In T4 the user needs to control for multiple variables. For example, an expert tests the hypothesis that FA1 is reduced in diseased subjects. We made sure all tasks were possible in both tools and structured them with increasing difficulty:
\begin{description}
\itemsep0em
\item \textbf{T1: Interpreting a single cluster of a single subject.}\\
\textbf{a)} (For a given cluster) Which value is higher FA1 or FA2?\\
\textbf{b)} (For a given cluster) Where along the fiber bundle is FA1 the highest?\\
\textbf{c)} (For a given cluster) Is the standard deviation of FA1 rather large or small compared to that of Estimated Uncertainty?
    \item \textbf{T2: Comparing multiple clusters of a single subject.}\\
    \textbf{a)} (For a given subject) Which are the two clusters with the highest estimated uncertainty?\\
    \textbf{b)} (For two given clusters and a given subject) Which cluster has a higher number of fibers in the bundle?
    \item \textbf{T3: Interaction between the same cluster in multiple subjects.}\\
        \textbf{a)} (For given subjects and a cluster) In which subject is FA1 maximal?\\
        \textbf{b)} (For given subjects and a cluster) For which pair of subjects is the difference in FA2 minimal?
    \item \textbf{T4: Comparing multiple clusters of multiple subjects.}\\
        \textbf{a)} In which female subject cluster is FA2 maximal?\\
        \textbf{b)} In which cluster is FA1 minimal?\\
        \textbf{c)} (For each two given clusters of two subjects) Does subject A or B have a lower FA1?
    \item \textbf{T5: General usability of components/features.}\\
        \textbf{a)} In a view of your choice, color the data by subjects.\\
        \textbf{b)} Can you find a U-shaped cluster in the 3D visualization?
\end{description} \paragraph{\textbf{Procedure.}} Due to COVID19, the study was conducted online via 50-60 minutes arranged video conferences with the participants. We asked the participant to share their screens while working on the tasks. We assigned participants randomly to one of the two tools to avoid user bias. Each study session started with an introduction, demonstrating underlying visualization components with the fundamental interaction possibilities available in each tool. Then, users had 2 minutes to explore the tool and its main features, and were allowed to ask questions during this period. Then, we provided the participants with an online document describing the tasks to complete. 
Users wrote down short answers in the document after they thought each was done and immediately notified the experimenter when they finished a task. Users were not told if their answers were correct or wrong. The first task of each session (in addition to T1-5) served as an example to provide hands-on familiarity with the assigned tool. Following this training, the users performed the tasks while we measured their task-completion times. We budgeted a fixed timeframe of 5 minutes for each task. When a user was not able to solve a sub-task in this timeframe, we assigned a penalty of 150 seconds. After participants completed all tasks, they were asked to complete a post-study questionnaire accessing their demographic data and judgments of usability and qualitative feedback. Additionally, we used a standard NASA-TLX survey to assess the workload with 6 questions~\cite{hart1988development}.

\paragraph{\textbf{Expert study design.}} 
We recruited 6 domain experts to evaluate the performance of both AFQ-Browser and FiberStars. We randomly assigned half of the experts to the AFQ-Browser and the other half to  FiberStars. First, experts performed the same training task as the novices. After, we asked the participants to solve Task 1-4 with a twenty-minute overall time limit. For the expert study, we did not include Task 5 and instead asked for more extensive and detailed qualitative post-experiment feedback.

\section{Results}

The results of our user study show an advantage of FiberStars in terms of accuracy and significant improvements of processing time for exploring the data.

\subsection{Quantitative Analysis}

We conducted a quantitative statistical analysis by analyzing the answers of the $N=20$ novices and $N=6$ domain experts.

\paragraph{\textbf{Accuracy and Performance.}}

Regarding accuracy in exploring the given DTI data, we examine the correctness of results by verifying the user responses for each task. For a correct answer in a sub-task, we assigned 1 point, and for a false answer, we assigned 0 points. Overall, the 10 participants using AFQ-Browser answered, on average, $69.17\%$ of the tasks correctly ($SD=31.75\%$). The 10 non-expert participants using FiberStars were able to answer $87.5\%$ correctly ($SD=10.55\%$). 

\begin{table}[h]
	\centering
	\caption{Error rates among novice participants per tasks.}
	\def\arraystretch{1.15}
	\resizebox{\linewidth}{!}{    
	\begin{tabular}{lccccccccccccc}
    \hline
    \textbf{Task}&\textbf{1a}&\textbf{1b}&\textbf{1c}&\textbf{2a}&\textbf{2b}&\textbf{3a}&\textbf{3b}&\textbf{4a}&\textbf{4b}&\textbf{4c}&\textbf{5a}&\textbf{5b} & \textbf{Mean} \\
    \hline
    \textbf{AFQ} & 0 & \textbf{0} & 0.8 & 0.9 & 0.1 & \textbf{0} & 0.2 & \textbf{0.2} & 0.1 & 0.3 & 0.5 & 0.6 & 0.308 \\
    \textbf{FS}  & \textbf{0} & 0.2 & \textbf{0.3} & \textbf{0.2} & 0.1 & 0.1 & \textbf{0.1} & 0.3 & 0.1 & \textbf{0} & \textbf{0} & \textbf{0.1} & \textbf{0.125} \\
	\hline
    \end{tabular}
	}
	\label{tab:correctness_novices}
\end{table}


Comparing the correct answers in both tools, the AFQ users were able to answer question 1b and 3a more often correctly than FiberStars users. In all other sub-tasks, the users either performed equal or more accurately with FiberStars, as depicted by the error rates in Table~\ref{tab:correctness_novices}. For testing the significance in mean differences for both tools, we used a two-sided statistical t-test with the null hypothesis, assuming that the two independent group means for AFQ and FiberStars are equal. From these values, we received a p-value of 0.0354 ($p\leq0.05$). With respect to \textbf{H2}, we therefore conclude that novices performed significantly better on our tasks with FiberStars.

\paragraph{\textbf{Efficiency.}}
\begin{figure}[t]
  \centering
  \includegraphics[width=1\linewidth]{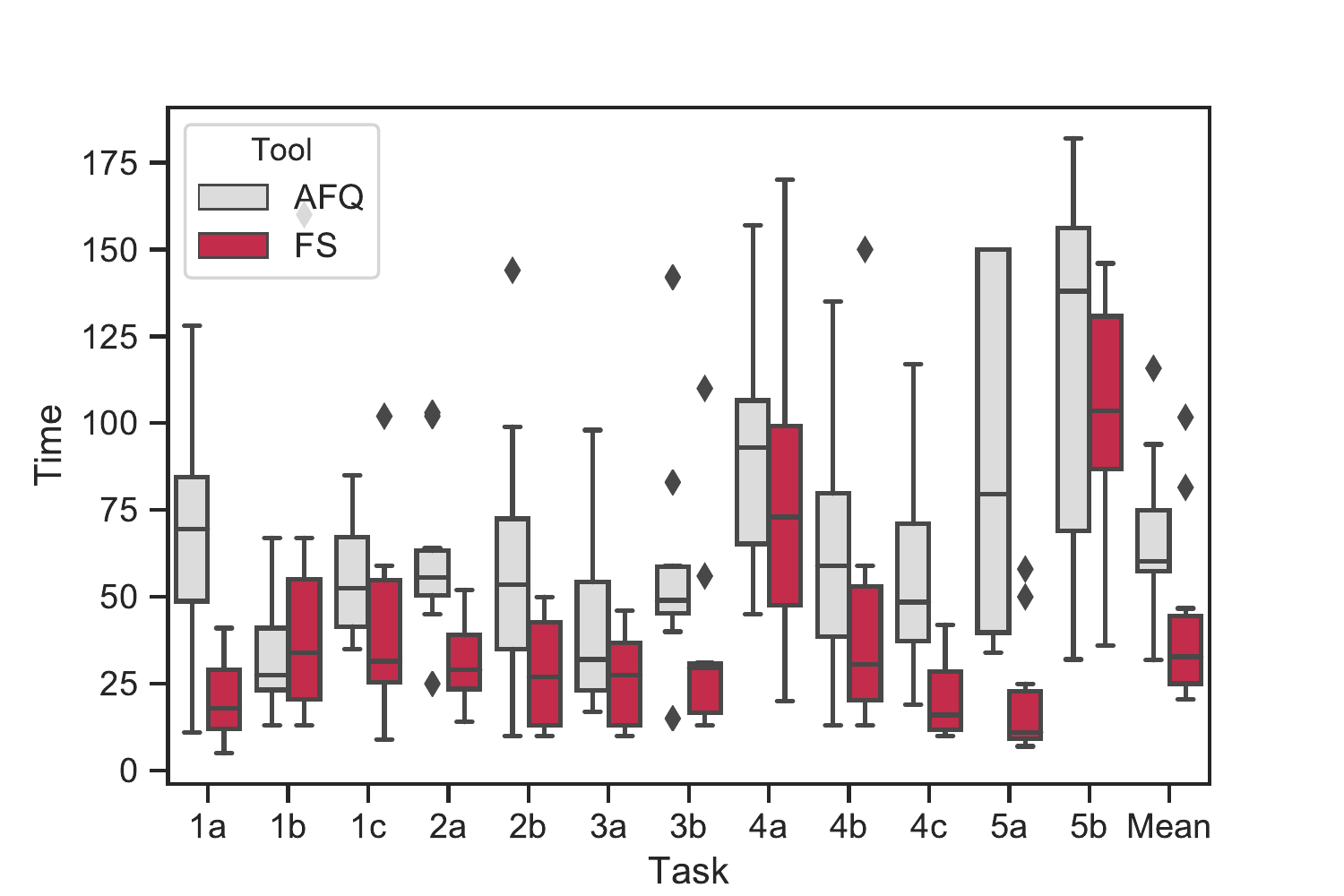}
  \caption{\textbf{Distribution of times per task for novices.} Times the users needed to solve the tasks for AFQ-Browser in grey (AFQ) and FiberStars in red (FS), across single sub-tasks. The right boxplots shows the average time over all tasks. Participants using FiberStars were faster in all cases. This difference is statistically significant.}
  \label{fig:meantimes}
\end{figure}

\begin{figure}
  \centering
  \includegraphics[width=1\linewidth]{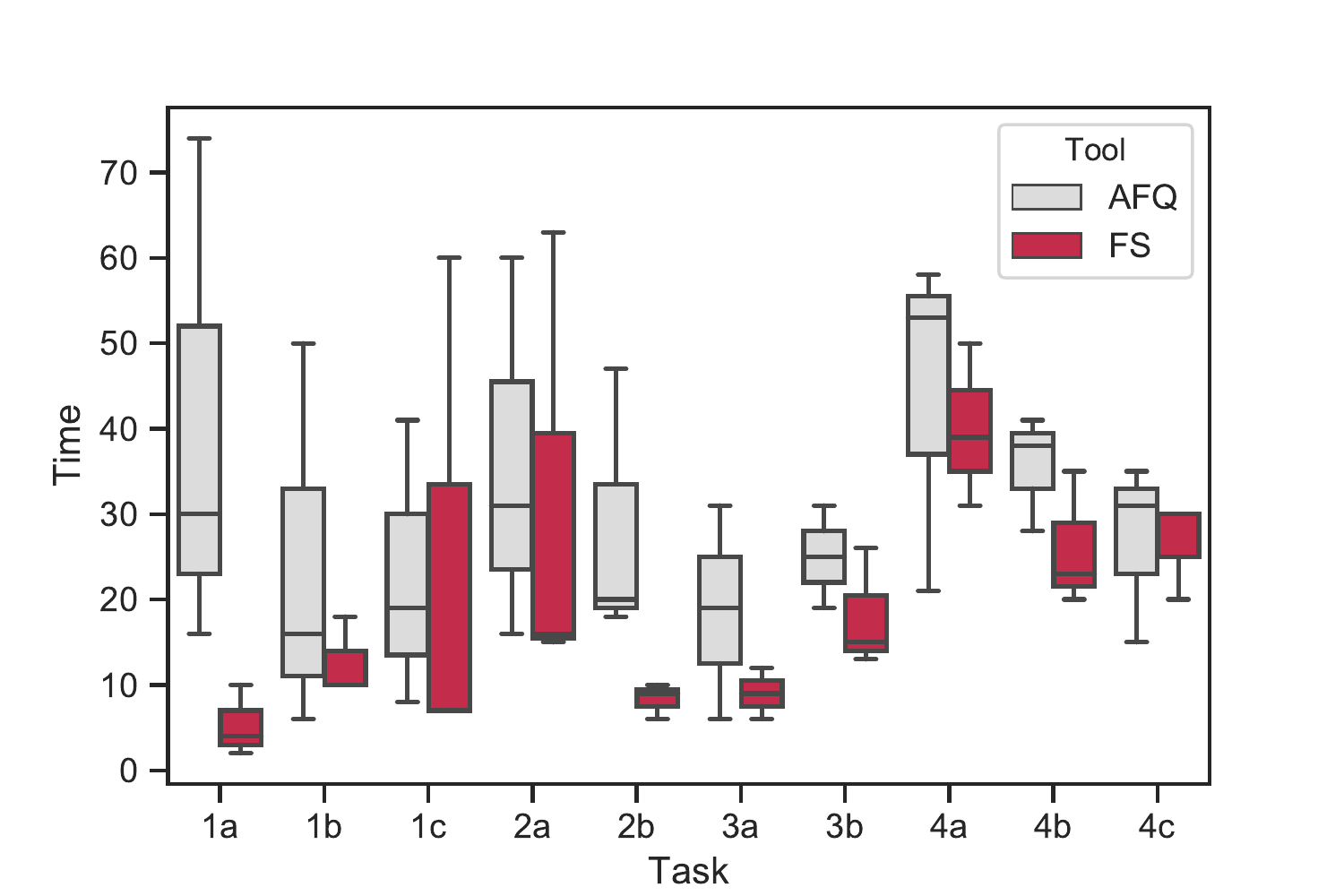}
  \caption{\textbf{Distribution of time needed per sub-task for experts.} Time the experts took to solve single sub-tasks grouped by tool where AFQ-Browser shown in grey (AFQ) and FiberStars in red (FS). This difference is statistically significant.}
  \label{fig:meantimes_experts}
\end{figure}

To test H3, we compared times the users needed to solve the tasks. 
The mean time per sub-task for the 10 AFQ users was 66.49 seconds ($SD=23.29$), and for FiberStars 41.40 seconds ($SD=25.41$). On average, FiberStars yields an improvement in speed of more than 37\% for non-experts. Task 1b took least average time with AFQ, while with FiberStars, users required the least time for Task 5a with 20.5 seconds. Figure~\ref{fig:meantimes} shows the distributions. Novices performed faster with FiberStars in all tasks except subtask 1b.
Summarizing all sub-tasks from Task 1 involving single subject single cluster problems, non-experts required 108.5 seconds on average with FiberStars, and 156.6 seconds with the alternative tool. Task 2 concerning multiple clusters of one subject, was solved in 121.6 with AFQ, and 58.9 seconds with FiberStars. Task 3 (one cluster across multiple subjects) required 100.6 seconds with AFQ, and 61 seconds with FiberStars. Task 4 comparing multiple subjects with multiple clusters, where AFQ users needed 212.2 seconds and FiberStars users 146.3 seconds. Task 5 took users 206.9 seconds with AFQ and 122.2 seconds with FiberStars. In general, users performed slowest in both tools when comparing multiple clusters across multiple subjects (high-level task). Overall, the expert users were faster with FiberStars in every task except for Task 1c. Improvement for domain experts was around 32\%, with an average of 30.1s for one subtask with AFQ and 20.2s for a subtask with Fiberstars. We state the null hypothesis that group means are equal without variation from both groups. The alternative hypothesis states differences between the group means. The resulting p-value is statistically significant with 0.0195 ($p\leq0.02$). We reject the null hypothesis in favor of the initial hypothesis \textbf{H3} with significant differences among both user groups. Overall, novices perform faster with our software.

\paragraph{\textbf{Usability.}}

\begin{table}
    \tiny
	\centering
	\caption{\textbf{Mean values of subjective responses of novices} showing the statements rated on a 7-point Likert-Scale (1 = totally disagree, 7 = totally agree). The asterisk * signs the statistically significant result.}
	\def\arraystretch{1.3}
	\resizebox{\linewidth}{!}{    
	\begin{tabular}{lrr}
	 \textbf{Statement}& \textbf{AFQ} & \textbf{FiberStars} \\
	 \midrule
	\textbf{The usability was very good}.& 5.8 $\pm 1.14$ & \textbf{6.2} $\pm 0.79$\\
	 
	 \textbf{I liked working with this tool.}& 6.0 $\pm 1.15$ & \textbf{6.4} $\pm 0.70$ \\
	 
	 \textbf{\vtop{\hbox{\strut I found it easy to compare}\hbox{\strut the data among subjects.}}}& 5.2 $\pm 1.81$& \textbf{5.8} $\pm 1.14$\\
	 
	\textbf{\vtop{\hbox{\strut I found it easy to navigate through}\hbox{\strut the different fiber clusters.}}} & 5.5 $\pm 1.18$ & \textbf{5.7} $\pm 0.82$\\
	
	\textbf{\vtop{\hbox{\strut The 2D plots or diagrams }\hbox{\strut were understandable.*}}}& 4.6 $\pm 1.58$& \textbf{5.7} $\pm 0.95$ \\
	
	\textbf{\vtop{\hbox{\strut The visualization of the }\hbox{\strut 3D fiber clusters were very pleasing.}}}& 6.3 $\pm 0.82$ & \textbf{6.8} $\pm 0.42$\\
	
	\textbf{\vtop{\hbox{\strut Additional information beside }\hbox{\strut the 2D plots were very helpful.*}}}& 5.1 $\pm 1.66$ & \textbf{6.4} $\pm 0.97$\\
 \hline
	\end{tabular}}
	\label{tab:usability}
\end{table}
We analyzed all non-expert answers ($N=20$). The Likert Scale questionnaire helps validating our findings, clarifying hypotheses as well as giving feedback to the usability results. The NASA-TLX questionnaire to access workload in terms of mental demand, frustration, effort, etc. did not yield any interesting results. By using participants' subjective questionnaire responses, we evaluate the perceived performance of both tools. The participant's responses were recorded on a 7 point Likert scale with 1 = totally disagree, and 7 = totally agree in a post-experimental questionnaire. FiberStars scored with a modest advantage compared to AFQ-Browser in perceived usability. Questions are stated in Table~\ref{tab:usability}. There was no statistical significance for these questions between both groups AFQ and FiberStars. Users rated FiberStars' usability on average with 6.2 ($SD=0.79$) and AFQ with 5.8 ($SD=1.14$) on the 1-7 scale. We could find a significant difference for the question \textit{'Additional information beside the 2D plots was very helpful.'} with $p=0.05$. Additional information in FiberStars was rated on average higher with 6.4 ($SD=0.97$) than AFQ with 5.1 ($SD=1.66$). Regarding \textbf{H1}, novices liked working with both tools but preferred FiberStars' 2D visualizations. During the study, we additionally recorded which FiberStars component was used to solve a given task. For Task 1, $43.3\%$ of the participants used the Comparison Matrix, whereas $56.6\%$ used the split-screen view. In Task 2, roughly $75\%$ preferred the split-screen view, while the other $25\%$ used the Matrix Comparison View. Task 3 was mainly solved with the split-screen view. 
Task 4 was dominantly solved with the Projection View ($53.33\%$), followed by the split-screen view $26.67\%$ and with the Matrix Comparison View by $20\%$. Task 5 was solved by $45\%$ with the Projection View and $55\%$ with the split-screen view. 
The use of a specific component depends on the given task. Tasks involving the comparison of different scalar values among a single subject or a single cluster were likely to be solved with either the split-screen view or the Matrix Comparison. For all tasks, the users found the answers in the radar charts. Tasks where users had to find values among a group with many clusters or subjects, they  mainly used the Projection View. Grouping values from lowest to highest was the most popular ordering choice there.
\paragraph{\textbf{Domain Expert performance.}}
The 3 experts testing AFQ-Browser needed, on average, 30.1 seconds per sub-task $SD=8.27$, while experts using FiberStars required 20.2 seconds to solve a sub-task $SD=11.35$. The difference in mean time for both AFQ and FiberStars, is statistically significant at 0.0200 ($p\leq0.05$) (\textbf{H3}). The distribution of how experts performed in terms of timing is shown in Figure~\ref{fig:meantimes_experts}. We used a two-sided t-test for testing statistical significance between novices and experts. Experts performed significantly faster than novices in both tools.
Furthermore, we could not find significant differences between both expert groups in terms of subjectively perceived usability (\textbf{H1}). AFQ and FiberStars experts perceived usability as good in both tools. In terms of accuracy, the experts using AFQ were able to answer $73.33\%$ ($SD=30.63\%$) of the tasks correctly on average, whereas the experts using FiberStars gave $96.67\%$ ($SD=0.11$) correct answers. We could find statistical significance between both group means in terms of correct answers for domain experts. 
With a p-value of around 0.0351, we confirm that there are significant differences among both groups in terms of correctness and 
reject the null hypothesis in favor of \textbf{H2}. For novices and experts, we could not find statistical significance in mean differences between the groups using AFQ and Fiberstars.
\begin{table}[h]
	\centering
	\caption{Error rates among expert participants per tasks.}
	\def\arraystretch{1.15}
	\resizebox{\linewidth}{!}{    
	\begin{tabular}{lccccccccccc}
    \hline
    \textbf{Task}&\textbf{1a}&\textbf{1b}&\textbf{1c}&\textbf{2a}&\textbf{2b}&\textbf{3a}&\textbf{3b}&\textbf{4a}&\textbf{4b}&\textbf{4c}& \textbf{Mean} \\
    \hline
    \textbf{AFQ} & 0 & 0.33 & 0.33 & 1 & 0 & 0 & 0 & 0.33 & 0.33 & 0.33 & 0.267 \\
    \textbf{FS}  & 0 & \textbf{0} & \textbf{0} & \textbf{0} & 0 & 0 & 0 & \textbf{0} & 0.33 & \textbf{0} & \textbf{0.033} \\
	\hline
    \end{tabular}
	}
	\label{tab:correctness_experts}
\end{table}


\subsection{Qualitative Results and Expert Feedback}

We collected useful qualitative feedback from the experts.
Most expert users had prior exposure to slicerDMRI~\cite{norton2017slicerdmri} or TrackVis~\cite{wang2007trackvis}, they were able to make themselves familiar quickly with both tools. In AFQ-Browser, users appreciated the brushing technique, which highlights clusters directly in the brain model. They further found the plots to be a good summary, despite having difficulties in differentiating single subjects. Experts missed a way to study exact values and would prefer more detailed statistics (mean or standard deviation for scalars). When comparing multiple clusters at once, participants requested an option to sort the 2D plots by scalar metrics in AFQ. Similarly, experts missed this sorting feature in the Comparison Matrix in FiberStars, despite having the option to sort by cluster or subject. However, they found it valuable to compare multiple numbers of clusters and subjects in the given level of detail. According to the users, the radar charts even reflected all of the essential information on DTI data. Experts would have liked to be able to configure the scalars shown in the radar chart. In the 3D visualization, a legend describing the color map would have been useful. The Universal Toolbar proved to be compelling, granting easy access to the data at all times. One expert expressed interest in FiberStars' compatibility with certain file formats. Summing up, multiple pros and cons in both tools underline our result from the previous section.

\section{Discussion}

Qualitative feedback and quantitative analysis indicate novices and experts appreciated usability of both, FiberStars and AFQ. All groups were able to quickly adapt to user-interface design and understand functionalities of each tool. Both cohorts were significantly faster completing the tasks in FiberStars as compared to AFQ. 
Unfortunately accuracy in AFQ did not significantly increase for experts. In FiberStars, however, observed significant improvement from novice to expert efficiency suggests experts are enabled to harvest more of their full potential. This is an essential finding, providing experts with a tool for accelerating the process of analyzing large scale data sets. Based on questionnaire findings, FiberStars' 2D plots were found to be significantly more helpful than AFQs'. Displaying metadata directly next to visualizations like 3D or radar charts is useful, but users still appreciate traditional tabular views as in AFQ. 
FiberStars' higher efficiency could stem from insights at the individual subject level combined with comparisons of multiple subjects. It was more difficult for non-experts to compare values across two subjects and two clusters in AFQ-Browser. This is reconfirmed by experts feeling challenged when analyzing multiple line plots while extracting and comparing data for several individual subjects. Moreover, participants face difficulties distinguishing subjects in the plots, as coloring by subject is not easily accessible in AFQ, which was also independently suggested by two experts in the qualitative feedback.
Participants using FiberStars preferred to consult the Projection View over the Comparison Matrix to compare multiple subjects and multiple clusters simultaneously to find overall trends or correlations in the data. We assume that the Projection View can have a significant impact in contributing to our initial goal in comparing multiple subjects or drilling down to a representative sample of interest, thereby facilitating the exploration process for domain experts. 

\paragraph{\textbf{Limitations.}}
Representing fiber bundle values in simple line plots is straight-forward and generally comprehensible. In FiberStars, we experimented with encoding values in tract color gradients directly as part of the 3-D rendering, which seems less favorable in some situations  (see Task 1b).

When working with users, we typically did not show more than 15 radial plots on-screen simultaneously. While shrinking or scrolling would allow representing even more components, we do not know at which level the otherwise useful comparison matrix would degenerate as fewer details can be spotted. However, previous small multiples visualizations frequently scale to in the order of 10 by 10 at a time~\cite{lekschas, liu2018correlatedmultiples, keefe}. We also want to emphasize that appropriate filtering steps (e.g., projection view Fig.~\ref{fig:teaser}) can lower demand for ever-larger matrix scaling. Moreover, with multi-subject DTI visualization still in its early attempts, it would have been desirable to evaluate FiberStars in the context of further alternatives, such as Fiber Model (DiffRadar).

\section{Conclusions and Future Work}


We have presented \textit{FiberStars}, a new open-source web-based visualization software to view extensive diffusion MRI datasets. In an iterative design process, we have derived requirements for analyzing such high-dimensional and longitudinal neuroscience datasets in web-based and scalable visualization software. Our resulting tool now supports the exploration of multiple fiber clusters across multiple subjects. The performed user study shows that it better supports experts on the given tasks and even lets novices gain insights from tractography data efficiently. In the future, we would like to investigate how we can automate steps in the analysis and create intelligent tractography exploration techniques taking even more data into account. Our findings and the open nature of our research will hopefully spur the adoption of web-based scientific visualizations and encourage further research in comparative visualizations and multidimensional scaling beyond the neurosciences.

\acknowledgments{
We would like to thank the participants of the user study and the authors of AFQ-Browser for helping with the installation and setup.}

\bibliographystyle{abbrv-doi}

\bibliography{template}

\begin{thebibliography}{10}

\bibitem{abbasloo2019interactive}
A.~Abbasloo, V.~Wiens, T.~Schmidt-Wilcke, P.~C. Sundgren, R.~Klein, and
  T.~Schultz.
\newblock Interactive formation of statistical hypotheses in diffusion tensor
  imaging.
\newblock In {\em VCBM}, pp. 33--43, 2019.

\bibitem{al2014neurolines}
A.~K. Al-Awami, J.~Beyer, H.~Strobelt, N.~Kasthuri, J.~W. Lichtman, H.~Pfister,
  and M.~Hadwiger.
\newblock Neurolines: a subway map metaphor for visualizing nanoscale neuronal
  connectivity.
\newblock {\em IEEE Transactions on Visualization and Computer Graphics},
  20(12):2369--2378, 2014.

\bibitem{assaf2008diffusion}
Y.~Assaf and O.~Pasternak.
\newblock Diffusion tensor imaging (dti)-based white matter mapping in brain
  research: a review.
\newblock {\em Journal of molecular neuroscience}, 34(1):51--61, 2008.

\bibitem{avram2016clinical}
A.~V. Avram, J.~E. Sarlls, A.~S. Barnett, E.~{\"O}zarslan, C.~Thomas, M.~O.
  Irfanoglu, E.~Hutchinson, C.~Pierpaoli, and P.~J. Basser.
\newblock Clinical feasibility of using mean apparent propagator (map) mri to
  characterize brain tissue microstructure.
\newblock {\em NeuroImage}, 127:422--434, 2016.

\bibitem{basser1994estimation}
P.~J. Basser, J.~Mattiello, and D.~LeBihan.
\newblock Estimation of the effective self-diffusion tensor from the nmr spin
  echo.
\newblock {\em Journal of Magnetic Resonance, Series B}, 103(3):247--254, 1994.

\bibitem{basser2000vivo}
P.~J. Basser, S.~Pajevic, C.~Pierpaoli, J.~Duda, and A.~Aldroubi.
\newblock In vivo fiber tractography using dt-mri data.
\newblock {\em Magnetic resonance in medicine}, 44(4):625--632, 2000.

\bibitem{basser2011microstructural}
P.~J. Basser and C.~Pierpaoli.
\newblock Microstructural and physiological features of tissues elucidated by
  quantitative-diffusion-tensor mri.
\newblock {\em Journal of magnetic resonance}, 213(2):560--570, 2011.

\bibitem{borkin2011evaluation}
M.~Borkin, K.~Gajos, A.~Peters, D.~Mitsouras, S.~Melchionna, F.~Rybicki,
  C.~Feldman, and H.~Pfister.
\newblock Evaluation of artery visualizations for heart disease diagnosis.
\newblock {\em IEEE transactions on visualization and computer graphics},
  17(12):2479--2488, 2011.

\bibitem{bostock2011d3}
M.~Bostock, V.~Ogievetsky, and J.~Heer.
\newblock D$^3$ data-driven documents.
\newblock {\em IEEE transactions on visualization and computer graphics},
  17(12):2301--2309, 2011.

\bibitem{brandes2006eigensolver}
U.~Brandes and C.~Pich.
\newblock Eigensolver methods for progressive multidimensional scaling of large
  data.
\newblock In {\em International Symposium on Graph Drawing}, pp. 42--53.
  Springer, 2006.

\bibitem{brandes2008experimental}
U.~Brandes and C.~Pich.
\newblock An experimental study on distance-based graph drawing.
\newblock In {\em International Symposium on Graph Drawing}, pp. 218--229.
  Springer, 2008.

\bibitem{brehmer2013multi}
M.~Brehmer and T.~Munzner.
\newblock A multi-level typology of abstract visualization tasks.
\newblock {\em IEEE transactions on visualization and computer graphics},
  19(12):2376--2385, 2013.

\bibitem{cauteruccio2015automated}
F.~Cauteruccio, C.~Stamile, G.~Terracina, D.~Ursino, and D.~Sappey-Mariniery.
\newblock An automated string-based approach to white matter fiber-bundles
  clustering.
\newblock In {\em 2015 International Joint Conference on Neural Networks
  (IJCNN)}, pp. 1--8. IEEE, 2015.

\bibitem{chen2015fiber}
P.~Chen, X.~Fan, R.~Liu, X.~Tang, and H.~Cheng.
\newblock Fiber segmentation using a density-peaks clustering algorithm.
\newblock In {\em 2015 IEEE 12th International Symposium on Biomedical Imaging
  (ISBI)}, pp. 633--637. IEEE, 2015.

\bibitem{chen2008abstractive}
W.~Chen, S.~Zhang, S.~Correia, and D.~S. Ebert.
\newblock Abstractive representation and exploration of hierarchically
  clustered diffusion tensor fiber tracts.
\newblock In {\em Computer Graphics Forum}, vol.~27, pp. 1071--1078. Wiley
  Online Library, 2008.

\bibitem{chen2009novel}
W.~Chen, S.~Zhang, A.~MacKay-Brandt, S.~Correia, H.~Qu, J.~A. Crow, D.~F. Tate,
  Z.~Yan, Q.~Peng, et~al.
\newblock A novel interface for interactive exploration of dti fibers.
\newblock {\em IEEE Transactions on Visualization and Computer Graphics},
  15(6):1433--1440, 2009.

\bibitem{conturo1999tracking}
T.~E. Conturo, N.~F. Lori, T.~S. Cull, E.~Akbudak, A.~Z. Snyder, J.~S. Shimony,
  R.~C. McKinstry, H.~Burton, and M.~E. Raichle.
\newblock Tracking neuronal fiber pathways in the living human brain.
\newblock {\em Proceedings of the National Academy of Sciences},
  96(18):10422--10427, 1999.

\bibitem{de2004sparse}
V.~De~Silva and J.~B. Tenenbaum.
\newblock Sparse multidimensional scaling using landmark points.
\newblock Technical report, Technical report, Stanford University, 2004.

\bibitem{dibiase}
M.~A. Di~Biase, F.~Zhang, A.~Lyall, M.~Kubicki, R.~C.~W. Mandl, I.~E. Sommer,
  and O.~Pasternak.
\newblock Neuroimaging auditory verbal hallucinations in schizophrenia patient
  and healthy populations.
\newblock {\em Psychological Medicine}, 50(3):403–412, 2020. doi: {{%
10\hspace{.1pt}\discretionary{.}{%
}{.}\hspace{.4pt}1017\discretionary{/}{%
}{/}S0033291719000205}}


\bibitem{draper2009survey}
G.~M. Draper, Y.~Livnat, and R.~F. Riesenfeld.
\newblock A survey of radial methods for information visualization.
\newblock {\em IEEE transactions on visualization and computer graphics},
  15(5):759--776, 2009.

\bibitem{faulkner2003beyond}
L.~Faulkner.
\newblock Beyond the five-user assumption: Benefits of increased sample sizes
  in usability testing.
\newblock {\em Behavior Research Methods, Instruments, \& Computers},
  35(3):379--383, 2003.

\bibitem{fornito2015connectomics}
A.~Fornito, A.~Zalesky, and M.~Breakspear.
\newblock The connectomics of brain disorders.
\newblock {\em Nature Reviews Neuroscience}, 16(3):159, 2015.

\bibitem{gori2016parsimonious}
P.~Gori, O.~Colliot, L.~Marrakchi-Kacem, Y.~Worbe, F.~D.~V. Fallani, M.~Chavez,
  C.~Poupon, A.~Hartmann, N.~Ayache, and S.~Durrleman.
\newblock Parsimonious approximation of streamline trajectories in white matter
  fiber bundles.
\newblock {\em IEEE transactions on medical imaging}, 35(12):2609--2619, 2016.

\bibitem{haehn2020trako}
D.~Haehn, L.~Franke, F.~Zhang, S.~Cetin-Karayumak, S.~Pieper, L.~J. O'Donnell,
  and Y.~Rathi.
\newblock Trako: Efficient transmission of tractography data for visualization.
\newblock In {\em Medical Image Computing and Computer Assisted Intervention --
  MICCAI 2020}, pp. 322--332. Springer International Publishing, Cham, 2020.

\bibitem{haehn2014neuroimaging}
D.~Haehn, N.~Rannou, B.~Ahtam, E.~Grant, and R.~Pienaar.
\newblock Neuroimaging in the browser using the x toolkit.
\newblock {\em Frontiers in Neuroinformatics}, 101, 2014.

\bibitem{hart1988development}
S.~G. Hart and L.~E. Staveland.
\newblock Development of nasa-tlx (task load index): Results of empirical and
  theoretical research.
\newblock In {\em Advances in psychology}, vol.~52, pp. 139--183. Elsevier,
  1988.

\bibitem{hwang2010number}
W.~Hwang and G.~Salvendy.
\newblock Number of people required for usability evaluation: the 10$\pm$2
  rule.
\newblock {\em Communications of the ACM}, 53(5):130--133, 2010.

\bibitem{jernigan2018adolescent}
T.~L. Jernigan, S.~A. Brown, and G.~J. Dowling.
\newblock The adolescent brain cognitive development study.
\newblock {\em Journal of research on adolescence: the official journal of the
  Society for Research on Adolescence}, 28(1):154--156, 2018.

\bibitem{jianu2009exploring}
R.~Jianu, C.~Demiralp, and D.~Laidlaw.
\newblock Exploring 3d dti fiber tracts with linked 2d representations.
\newblock {\em IEEE transactions on visualization and computer graphics},
  15(6):1449--1456, 2009.

\bibitem{jianu2011exploring}
R.~Jianu, C.~Demiralp, and D.~H. Laidlaw.
\newblock Exploring brain connectivity with two-dimensional neural maps.
\newblock {\em IEEE transactions on visualization and computer graphics},
  18(6):978--987, 2011.

\bibitem{jonsson2019visual}
D.~J{\"o}nsson, A.~Bergstr{\"o}m, C.~Forsell, R.~Simon, M.~Engstr{\"o}m,
  A.~Ynnerman, and I.~Hotz.
\newblock A visual environment for hypothesis formation and reasoning in
  studies with fmri and multivariate clinical data.
\newblock In {\em Eurographics Workshop on Visual Computing for Biology and
  Medicine}, 2019.

\bibitem{kamali2016automated}
T.~Kamali and D.~Stashuk.
\newblock Automated segmentation of white matter fiber bundles using diffusion
  tensor imaging data and a new density based clustering algorithm.
\newblock {\em Artificial intelligence in medicine}, 73:14--22, 2016.

\bibitem{keefe}
D.~{Keefe}, M.~{Ewert}, W.~{Ribarsky}, and R.~{Chang}.
\newblock Interactive coordinated multiple-view visualization of biomechanical
  motion data.
\newblock {\em IEEE Transactions on Visualization and Computer Graphics},
  15(6):1383--1390, 2009. doi: {{%
10\hspace{.1pt}\discretionary{.}{%
}{.}\hspace{.4pt}1109\discretionary{/}{%
}{/}TVCG\hspace{.1pt}\discretionary{.}{%
}{.}\hspace{.4pt}2009\hspace{.1pt}\discretionary{.}{%
}{.}\hspace{.4pt}152}}


\bibitem{khatami2017bundlemap}
M.~Khatami, T.~Schmidt-Wilcke, P.~C. Sundgren, A.~Abbasloo, B.~Sch{\"o}lkopf,
  and T.~Schultz.
\newblock Bundlemap: Anatomically localized classification, regression, and
  hypothesis testing in diffusion mri.
\newblock {\em Pattern Recognition}, 63:593--600, 2017.

\bibitem{ledoux2017fiberweb}
L.-P. Ledoux, F.~C. Morency, M.~Cousineau, J.-C. Houde, K.~Whittingstall, and
  M.~Descoteaux.
\newblock Fiberweb: diffusion visualization and processing in the browser.
\newblock {\em Frontiers in neuroinformatics}, 11:54, 2017.

\bibitem{lekschas}
F.~{Lekschas}, B.~{Bach}, P.~{Kerpedjiev}, N.~{Gehlenborg}, and H.~{Pfister}.
\newblock Hipiler: Visual exploration of large genome interaction matrices with
  interactive small multiples.
\newblock {\em IEEE Transactions on Visualization and Computer Graphics},
  24(1):522--531, 2018. doi: {{%
10\hspace{.1pt}\discretionary{.}{%
}{.}\hspace{.4pt}1109\discretionary{/}{%
}{/}TVCG\hspace{.1pt}\discretionary{.}{%
}{.}\hspace{.4pt}2017\hspace{.1pt}\discretionary{.}{%
}{.}\hspace{.4pt}2745978}}


\bibitem{lichtman2011big}
J.~W. Lichtman and W.~Denk.
\newblock The big and the small: challenges of imaging the brain’s circuits.
\newblock {\em Science}, 334(6056):618--623, 2011.

\bibitem{liu2018correlatedmultiples}
X.~Liu, Y.~Hu, S.~North, and H.-W. Shen.
\newblock Correlatedmultiples: Spatially coherent small multiples with
  constrained multi-dimensional scaling.
\newblock In {\em Computer Graphics Forum}, vol.~37, pp. 7--18. Wiley Online
  Library, 2018.

\bibitem{maaten2008visualizing}
L.~v.~d. Maaten and G.~Hinton.
\newblock Visualizing data using t-sne.
\newblock {\em Journal of machine learning research}, 9(Nov):2579--2605, 2008.

\bibitem{5559623}
J.~G. {Malcolm}, M.~E. {Shenton}, and Y.~{Rathi}.
\newblock Filtered multitensor tractography.
\newblock {\em IEEE Transactions on Medical Imaging}, 29(9):1664--1675, 2010.

\bibitem{mcinnes2018umap}
L.~McInnes, J.~Healy, and J.~Melville.
\newblock Umap: Uniform manifold approximation and projection for dimension
  reduction.
\newblock {\em arXiv preprint arXiv:1802.03426}, 2018.

\bibitem{mei2016visually}
H.~Mei, H.~Chen, F.~Guo, F.~Zhang, W.~Chen, Z.~Song, and G.~Wang.
\newblock Visually exploring differences of dti fiber models.
\newblock In {\em International Conference on Technologies for E-Learning and
  Digital Entertainment}, pp. 333--344. Springer, 2016.

\bibitem{mohammed2017abstractocyte}
H.~Mohammed, A.~K. Al-Awami, J.~Beyer, C.~Cali, P.~Magistretti, H.~Pfister, and
  M.~Hadwiger.
\newblock Abstractocyte: a visual tool for exploring nanoscale astroglial
  cells.
\newblock {\em IEEE transactions on visualization and computer graphics},
  24(1):853--861, 2017.

\bibitem{nielsen1993mathematical}
J.~Nielsen and T.~K. Landauer.
\newblock A mathematical model of the finding of usability problems.
\newblock In {\em Proceedings of the INTERACT'93 and CHI'93 conference on Human
  factors in computing systems}, pp. 206--213, 1993.

\bibitem{norton2017slicerdmri}
I.~Norton, W.~I. Essayed, F.~Zhang, S.~Pujol, A.~Yarmarkovich, A.~J. Golby,
  G.~Kindlmann, D.~Wassermann, R.~S.~J. Estepar, Y.~Rathi, et~al.
\newblock Slicerdmri: open source diffusion mri software for brain cancer
  research.
\newblock {\em Cancer research}, 77(21):e101--e103, 2017.

\bibitem{o2015does}
L.~J. O'Donnell and O.~Pasternak.
\newblock Does diffusion mri tell us anything about the white matter? an
  overview of methods and pitfalls.
\newblock {\em Schizophrenia research}, 161(1):133--141, 2015.

\bibitem{pandey2019cerebrovis}
A.~Pandey, H.~Shukla, G.~S. Young, L.~Qin, A.~A. Zamani, L.~Hsu, R.~Huang,
  C.~Dunne, and M.~A. Borkin.
\newblock Cerebrovis: Designing an abstract yet spatially contextualized
  cerebral artery network visualization.
\newblock {\em IEEE transactions on visualization and computer graphics},
  26(1):938--948, 2019.

\bibitem{poco2012employing}
J.~Poco, D.~M. Eler, F.~V. Paulovich, and R.~Minghim.
\newblock Employing 2d projections for fast visual exploration of large fiber
  tracking data.
\newblock In {\em Computer Graphics Forum}, vol.~31, pp. 1075--1084. Wiley
  Online Library, 2012.

\bibitem{rathi2014multi}
Y.~Rathi, O.~Michailovich, F.~Laun, K.~Setsompop, P.~E. Grant, and C.-F.
  Westin.
\newblock Multi-shell diffusion signal recovery from sparse measurements.
\newblock {\em Medical image analysis}, 18(7):1143--1156, 2014.

\bibitem{rathi2011sparse}
Y.~Rathi, O.~Michailovich, K.~Setsompop, S.~Bouix, M.~E. Shenton, and C.-F.
  Westin.
\newblock Sparse multi-shell diffusion imaging.
\newblock In {\em International Conference on Medical Image Computing and
  Computer-Assisted Intervention}, pp. 58--65. Springer, 2011.

\bibitem{reddy2016joint}
C.~P. Reddy and Y.~Rathi.
\newblock Joint multi-fiber noddi parameter estimation and tractography using
  the unscented information filter.
\newblock {\em Frontiers in neuroscience}, 10:166, 2016.

\bibitem{saary2008radar}
M.~J. Saary.
\newblock Radar plots: a useful way for presenting multivariate health care
  data.
\newblock {\em Journal of clinical epidemiology}, 61(4):311--317, 2008.

\bibitem{schroeder2004visualization}
W.~J. Schroeder, B.~Lorensen, and K.~Martin.
\newblock {\em The visualization toolkit: an object-oriented approach to 3D
  graphics}.
\newblock Kitware, 2004.

\bibitem{thomason2011diffusion}
M.~E. Thomason and P.~M. Thompson.
\newblock Diffusion imaging, white matter, and psychopathology.
\newblock {\em Annual review of clinical psychology}, 7, 2011.

\bibitem{torgerson1952multidimensional}
W.~S. Torgerson.
\newblock Multidimensional scaling: I. theory and method.
\newblock {\em Psychometrika}, 17(4):401--419, 1952.

\bibitem{van2013wu}
D.~C. Van~Essen, S.~M. Smith, D.~M. Barch, T.~E. Behrens, E.~Yacoub,
  K.~Ugurbil, W.-M.~H. Consortium, et~al.
\newblock The wu-minn human connectome project: an overview.
\newblock {\em Neuroimage}, 80:62--79, 2013.

\bibitem{wang2007trackvis}
R.~Wang and V.~J. Wedeen.
\newblock Trackvis. org.
\newblock {\em Martinos Center for Biomedical Imaging, Massachusetts General
  Hospital}, 2007.

\bibitem{weidele2016graphical}
D.~Weidele, M.~van Garderen, M.~Golitko, G.~M. Feinman, and U.~Brandes.
\newblock On graphical representations of similarity in geo-temporal frequency
  data.
\newblock {\em Journal of Archaeological Science}, 72:105--116, 2016.

\bibitem{wu2019detecting}
W.~Wu, G.~McAnulty, H.~M. Hamoda, K.~Sarill, S.~Karmacharya, B.~Gagoski,
  L.~Ning, P.~E. Grant, M.~E. Shenton, D.~P. Waber, et~al.
\newblock Detecting microstructural white matter abnormalities of frontal
  pathways in children with adhd using advanced diffusion models.
\newblock {\em Brain imaging and behavior}, pp. 1--17, 2019.

\bibitem{yeatman2018browser}
J.~D. Yeatman, A.~Richie-Halford, J.~K. Smith, A.~Keshavan, and A.~Rokem.
\newblock A browser-based tool for visualization and analysis of diffusion mri
  data.
\newblock {\em Nature communications}, 9(1):1--10, 2018.

\bibitem{zhang2018suprathreshold}
F.~Zhang, W.~Wu, L.~Ning, G.~McAnulty, D.~Waber, B.~Gagoski, K.~Sarill, H.~M.
  Hamoda, Y.~Song, W.~Cai, et~al.
\newblock Suprathreshold fiber cluster statistics: Leveraging white matter
  geometry to enhance tractography statistical analysis.
\newblock {\em NeuroImage}, 171:341--354, 2018.

\bibitem{zhang2018anatomically}
F.~Zhang, Y.~Wu, I.~Norton, L.~Rigolo, Y.~Rathi, N.~Makris, and L.~J.
  O'Donnell.
\newblock An anatomically curated fiber clustering white matter atlas for
  consistent white matter tract parcellation across the lifespan.
\newblock {\em NeuroImage}, 179:429--447, 2018.

\end{thebibliography}
\end{document}